# On an extension of Lesche stability

A. El Kaabouchi[1], A. Le Méhauté[1], L. Nivanen[1], C.J. Ou[1,2] and A. Q. Wang[1]

[1]Institut Supérieur des Matériaux et Mécaniques Avancés du Mans
44 Avenue Bartholdi, 72000 Le Mans, France

[2]College of Information Science and Engineering,
Huaqiao University, Quanzhou 362021, China

**Abstract**

In this paper, we give a new method for proving the Lesche stability of several functionals (Incomplete entropy, Tsallis entropy, $\kappa-$entropy, Quantum-Group entropy). We prove also that the Incomplete $q-$expectation value and Rényi entropy for $(0<q<1)$ are $\alpha-$stable for all $(0<\alpha\leq q)$. Finally, we prove that the Incomplete $q-$expectation value is $\alpha-$stable for all $0<\alpha\leq 1$.

**Notations and definitions**

We denote by $E_f = \bigcup_{n \in \mathbf{N}^*} \mathbf{R}^n$. For $N \in \mathbf{N}^*$ and $\alpha > 0$, we define the function $d_{\alpha,N} : \mathbf{R}^N \times \mathbf{R}^N \to \mathbf{R}_+$ by

$$d_{\alpha,N}(p, p') = \left( \sum_{i=1}^{N} |p_i - p'_i|^\alpha \right)^{1/\alpha}.$$

We recall that

for $\alpha \geq 1$, $d_{\alpha,N}$ is a *distance* on $\mathbf{R}^N$

$$\forall p, p' \in \mathbf{R}^N, \ d_{\alpha,N}(p, p') = 0 \Leftrightarrow p = p'$$
$$\forall p, p' \in \mathbf{R}^N \ d_{\alpha,N}(p, p') = d_{\alpha,N}(p', p)$$
$$\forall p, p', p'' \in \mathbf{R}^N, \ d_{\alpha,N}(p, p'') \leq d_{\alpha,N}(p, p') + d_{\alpha,N}(p', p'')$$

for $\alpha < 1$, $d_{\alpha,N}$ is a *quasi-distance* on $\mathbf{R}^N$.

$$\forall p, p' \in \mathbf{R}^N, \ d_{\alpha,N}(p, p') = 0 \Leftrightarrow p = p'$$
$$\forall p, p' \in \mathbf{R}^N \ d_{\alpha,N}(p, p') = d_{\alpha,N}(p', p)$$
$$\forall p, p', p'' \in \mathbf{R}^N, \ d_{\alpha,N}(p, p'') \leq 2^{\frac{1}{\alpha}-1} \left( d_{\alpha,N}(p, p') + d_{\alpha,N}(p', p'') \right)$$

Let $C$ be a function defined on $A$ a subset of $E_f$. For $N \in \mathbf{N}^*$ and $\alpha > 0$, we denote by $C_{N,\max}$ the number defined by (when it exists) $C_{N,\max} = \sup\{|C(p)|, p \in A \cap \mathbf{R}^N\}$.

**Definition 1**

A Function $C$ defined on $A$ a subset of $E_f$ is said to be $\alpha - stable$ if $C$ posses the following property

$$\forall \varepsilon > 0, \ \exists \delta > 0, \ \forall N \in \mathbf{N}^*, \ \forall p, p' \in A \cap \mathbf{R}^N, \ d_{\alpha,N}(p, p') < \delta \Rightarrow \left| \frac{C(p) - C(p')}{C_{N,\max}} \right| < \varepsilon.$$

**Remark**

The Lesche stability [1] correspond to $\alpha - stability$ for $\alpha = 1$.

**Proposition 2**

Let $0 < \beta \leq \alpha$ and $C$ be a function defined on a subset of $E_f$.

If $C$ is $\alpha - stable$ then $C$ is $\beta - stable$.

*Proof.*

It's easy to see that for all $x \in ]0,1[$, $x^\alpha \leq x^\beta$ and thus if $d_{\beta,N}(p, p') < \delta$ then $d_{\alpha,N}(p, p') < \delta^{\beta/\alpha}$.

**Proposition 3**

If $C_1, C_2$ are $\alpha - stable$ with $C_1 \geq 0$, $C_2 \geq 0$ and $\lambda, \mu \in \mathbf{R}_+$, then $\lambda C_1 + \mu C_2$ is $\alpha - stable$.

*Proof.*

We can suppose that $C_1 \not\equiv 0$, $C_2 \not\equiv 0$ and $\lambda, \mu \in \mathbf{R}_+^*$.

We remark that for all $N \in \mathbf{N}^*$ and $\alpha > 0$,

$$\left| \frac{(\lambda C_1 + \mu C_2)(p) - (\lambda C_1 + \mu C_2)(p')}{(\lambda C_1 + \mu C_2)_{N,\max}} \right| = \frac{|\lambda(C_1(p) - C_1(p')) + \mu(C_2(p) - C_2(p'))|}{(\lambda C_1 + \mu C_2)_{N,\max}}$$

$$\leq \lambda \frac{|(C_1(p) - C_1(p'))|}{(\lambda C_1 + \mu C_2)_{N,\max}} + \mu \frac{|(C_2(p) - C_2(p'))|}{(\lambda C_1 + \mu C_2)_{N,\max}}$$

$$\leq \lambda \frac{|(C_1(p) - C_1(p'))|}{\lambda (C_1)_{N,\max}} + \mu \frac{|(C_2(p) - C_2(p'))|}{\mu (C_2)_{N,\max}}$$

$$= \frac{|(C_1(p) - C_1(p'))|}{(C_1)_{N,\max}} + \frac{|(C_2(p) - C_2(p'))|}{(C_2)_{N,\max}}$$

Let $\varepsilon > 0$, there exists $\delta_1 > 0$, $\delta_2 > 0$, such that $\forall N \in \mathbf{N}^*$, $\forall p, p' \in A \cap \mathbf{R}^N$,

$$d_{\alpha,N}(p,p') < \delta_1 \Rightarrow \left|\frac{C_1(p) - C_1(p')}{(C_1)_{N,\max}}\right| < \frac{\varepsilon}{2} \text{ and } d_{\alpha,N}(p,p') < \delta_2 \Rightarrow \left|\frac{C_2(p) - C_2(p')}{(C_2)_{N,\max}}\right| < \frac{\varepsilon}{2}.$$

We put $\delta = \min(\delta_1, \delta_2) > 0$, $\forall p, p' \in A \cap \mathbf{R}^N$,

$$d_{\alpha,N}(p,p') < \delta \Rightarrow \left(d_{\alpha,N}(p,p') < \delta_1 \text{ and } d_{\alpha,N}(p,p') < \delta_2\right)$$

Consequently $\left|\dfrac{(\lambda C_1 + \mu C_2)(p) - (\lambda C_1 + \mu C_2)(p')}{(\lambda C_1 + \mu C_2)_{N,\max}}\right| < \dfrac{\varepsilon}{2} + \dfrac{\varepsilon}{2} = \varepsilon$.

**Definition 4**

Let $q \in ]0,+\infty[\setminus\{1\}$. The incomplete entropy $S_q^I$ [2] is defined on $A = \bigcup_{n \in \mathbf{N}^*} \left\{p \in [0,1]^n, \sum_{i=1}^{n}(p_i)^q = 1\right\}$ by

$$S_q^I(p) = \frac{1 - \sum_{i=1}^{N} p_i}{1 - q} \text{ for all } p \in [0,1]^N.$$

**Theorem 5**

For all $q \in ]0,+\infty[\setminus\{1\}$, the incomplete entropy $S_q^I$ is $\alpha$-stable for all $\alpha \leq 1$.

*Proof.*

It's easy to see that

i) $\left|(S_q^I)_{N,\max}\right| = \left|S_q^I\left((1/N)^{1/q}, \cdots, (1/N)^{1/q}\right)\right| = \left|\dfrac{1 - N^{1-\frac{1}{q}}}{1 - q}\right|$.

ii) For all $q \in ]0,+\infty[\setminus\{1\}$, the function $x \mapsto \dfrac{1}{1 - x^{1-\frac{1}{q}}}$ is bounded on $[2,+\infty[$

Indeed if $q \in ]0,1[$, then $\lim_{x \to +\infty} \dfrac{1}{1 - x^{1-\frac{1}{q}}} = 1$ and if $q > 1$, then $\lim_{x \to +\infty} \dfrac{1}{1 - x^{1-\frac{1}{q}}} = 0$.

Let $M > 0$, such that for all $N \in \mathbf{N}^* \setminus \{1\}$, $\left|\dfrac{1}{1 - N^{1-\frac{1}{q}}}\right| \leq M$.

For all $\varepsilon > 0$, there exists $\delta = \left(\dfrac{\varepsilon}{M}\right)^{1/\alpha} > 0$, such that for all $N \in \mathbf{N}^*$, $p, p' \in A \cap \mathbf{R}^N$, $d_{\alpha,N}(p,p') < \delta$ implies

$$\left|\frac{S_q^I(p) - S_q^I(p')}{(S_q^I)_{N,\max}}\right| = \left|\frac{\sum_{i=1}^{N}(p_i' - p_i)}{1 - N^{1-\frac{1}{q}}}\right| \leq \frac{\sum_{i=1}^{N}|p_i' - p_i|}{\left|1 - N^{1-\frac{1}{q}}\right|}$$

$$\leq M \sum_{i=1}^{N}|p_i' - p_i| \leq M \sum_{i=1}^{N}|p_i' - p_i|^\alpha$$

$$= M\left(d_{\alpha,N}(p,p')\right)^\alpha < M\delta^\alpha = \varepsilon$$

**Definition 6**

Let $q \in ]0,+\infty[\setminus\{1\}$. The Tsallis entropy $S_q$ [3] is defined on $A = \bigcup_{n \in \mathbf{N}^*} \left\{p \in [0,1]^n, \sum_{i=1}^{n} p_i = 1\right\}$ by

$$S_q(p) = \frac{1 - \sum_{i=1}^{N}(p_i)^q}{q - 1} \text{ for all } p \in [0,1]^N.$$

**Theorem 7**
For all $q \in \left]0,+\infty\right[\setminus\{1\}$, the Tsallis entropy $S_q$ is $\alpha$-stable for all $\alpha \leq 1$.

**Corollary 8**
a) For all $\kappa \in \left]-1,1\right[$, the $\kappa$-entropy is $\alpha$-stable for all $\alpha \leq 1$.
b) For all $q \in \left]0,+\infty\right[\setminus\{1\}$, the Quantum-Group entropy [4] is $\alpha$-stable for all $\alpha \leq 1$.

*Proof.*

a) Let $\kappa \in \left]-1,1\right[$. The $\kappa$-entropy $S^\kappa$ is defined on $A = \bigcup_{n \in \mathbf{N}^*}\left\{p \in [0,1]^n, \sum_{i=1}^{n} p_i = 1\right\}$ by

$$S^\kappa(p) = -\frac{\sum_{i=1}^{N} p_i\left((p_i)^\kappa - (p_i)^{-\kappa}\right)}{2\kappa} \text{ for all } p \in [0,1]^N.$$

It's easy to see that $S^\kappa(p) = \frac{1}{2}\left(S_{\kappa+1}(p) + S_{1-\kappa}(p)\right)$ and we conclude by proposition 3.

b) Let $q \in \left]0,+\infty\right[\setminus\{1\}$. The Quantum-Group entropy is defined on $A = \bigcup_{n \in \mathbf{N}^*}\left\{p \in [0,1]^n, \sum_{i=1}^{n} p_i = 1\right\}$ is defined by

$$S_q^{QG}(p) = -\frac{\sum_{i=1}^{N}\left((p_i)^q - (p_i)^{1/q}\right)}{q - \frac{1}{q}} \text{ for all } p \in [0,1]^N.$$

It's easy to see that $S_q^{QG}(p) = \frac{q}{q+1}S_q(p) + \frac{\frac{1}{q}}{\frac{1}{q}+1}S_{\frac{1}{q}}(p) = \frac{q}{q+1}S_q(p) + \frac{1}{q+1}S_{\frac{1}{q}}(p)$ and we conclude by proposition 3.

For the proof of the theorem 7, we have need of the following lemmas.

**Lemma 9**
a) For all $q \in \left]0,1\right[$, the function $x \mapsto \frac{x^{1-q}}{1 - x^{1-q}}$ is bounded on $[2,+\infty[$ by $M_1 > 0$;

b) For all $q > 1$, the function $x \mapsto \frac{1}{1 - x^{1-q}}$ is bounded on $[2,+\infty[$ by $M_2 > 0$.

Indeed if $q \in \left]0,1\right[$, then $\lim_{x \to +\infty} \frac{x^{1-q}}{1-x^{1-q}} = 1$ and if $q > 1$, then $\lim_{x \to +\infty} \frac{1}{1-x^{1-q}} = 0$.

**Lemma 10**
Let $\alpha > 0$, $N \in \mathbf{N}^*$, $(p_i), (p_i') \in [0,1]^N$.

a) If $\alpha < 1$, then $\sum_{i=1}^{N}\left|p_i^\alpha - (p_i')^\alpha\right| \leq (d_{\alpha,N}(p,p'))^\alpha$;

b) If $\alpha < 1$, then $\sum_{i=1}^{N}\left|p_i^\alpha - (p_i')^\alpha\right| \leq N^{1-\alpha}(d_{1,N}(p,p'))^\alpha$;

c) If $\alpha > 1$, then $\sum_{i=1}^{N}\left|p_i^\alpha - (p_i')^\alpha\right| \leq \alpha(d_{1,N}(p,p'))$;

d) If $\alpha > 1$, then $\sum_{i=1}^{N}\left|p_i^\alpha - (p_i')^\alpha\right| \leq \alpha N^{1-\frac{1}{\alpha}}(d_{\alpha,N}(p,p'))$.

*Proof.*
a) We put $I_1 = \{i \in \{1,\cdots,N\}, p_i' < p_i\}$ and $I_2 = \{i \in \{1,\cdots,N\}, p_i < p_i'\}$.

For all $i \in I_1$, we have: $1 = \left(\dfrac{p_i'}{p_i}\right) + \left(1 - \dfrac{p_i'}{p_i}\right) \leq \left(\dfrac{p_i'}{p_i}\right)^\alpha + \left(1 - \dfrac{p_i'}{p_i}\right)^\alpha = \left(\dfrac{p_i'}{p_i}\right)^\alpha + \left|1 - \dfrac{p_i'}{p_i}\right|^\alpha$, we deduce that $(p_i)^\alpha - (p_i')^\alpha \leq |p_i - p_i'|^\alpha$. And by symmetry we deduce that all $i \in I_2$, we have: $(p_i')^\alpha - (p_i)^\alpha \leq |p_i - p_i'|^\alpha$.

We conclude that for $i \in \{1,\cdots,N\}$, $\left|(p_i)^\alpha - (p_i')^\alpha\right| \leq |p_i - p_i'|^\alpha$, and thus a) is proved.

b) By a) we have, $\sum_{i=1}^{N}\left|(p_i)^\alpha - (p_i')^\alpha\right| \leq \sum_{i=1}^{N}|p_i - p_i'|^\alpha = \sum_{i=1}^{N}|p_i - p_i'|^\alpha \cdot 1$, and by Holder's Inequality, $\left(\dfrac{1}{1/\alpha} + \dfrac{1}{1/(1-\alpha)} = 1\right)$, we obtain

$$\sum_{i=1}^{N}|p_i - p_i'|^\alpha \cdot 1 \leq \left(\sum_{i=1}^{N}\left(|p_i - p_i'|^\alpha\right)^{1/\alpha}\right)^\alpha \left(\sum_{i=1}^{N} 1^{1/(1-\alpha)}\right)^{1-\alpha} = \left(\sum_{i=1}^{N}|p_i - p_i'|\right)^\alpha N^{1-\alpha}.$$

c) The function $x \mapsto x^\alpha$ is continuous on $[0,1]$, differentiable on $]0,1[$, by the Mean-Value Theorem, we have

$$\forall i \in \{1,\cdots,N\}, \exists c_i \in (p_i, p_i'), \text{ such that } \left((p_i)^\alpha - (p_i')^\alpha\right) = (p_i - p_i')\alpha(c_i)^{\alpha-1},$$

We deduce that

$$\forall i \in \{1,\cdots,N\}, \left|(p_i)^\alpha - (p_i')^\alpha\right| \leq |p_i - p_i'|\alpha.$$

Consequently, $\sum_{i=1}^{N}\left|p_i^\alpha - (p_i')^\alpha\right| \leq \alpha\left(\sum_{i=1}^{N}|p_i - p_i'|\right) = \alpha d_{1,N}(p, p')$.

d) By c) we have $\sum_{i=1}^{N}\left|p_i^\alpha - (p_i')^\alpha\right| \leq \alpha\left(\sum_{i=1}^{N}|p_i - p_i'|\right) = \alpha\sum_{i=1}^{N}|p_i - p_i'| \cdot 1$, and by Holder's inequality we obtain

$$\sum_{i=1}^{N}|p_i - p_i'| \cdot 1 \leq \left(\sum_{i=1}^{N}|p_i - p_i'|^\alpha\right)^{1/\alpha}\left(\sum_{i=1}^{N} 1^{\alpha/(\alpha-1)}\right)^{1-\frac{1}{\alpha}} = N^{1-\frac{1}{\alpha}} d_{\alpha,N}(p, p').$$

Consequently, $\sum_{i=1}^{N}\left|p_i^\alpha - (p_i')^\alpha\right| \leq N^{1-\frac{1}{\alpha}} d_{\alpha,N}(p, p')$.

*Proof of theorem 7.*

It's easy to see that $\left|(S_q)_{\max}\right| = \left|S_q((1/N),\cdots,(1/N))\right| = \left|\dfrac{1 - N^{1-q}}{1-q}\right|$.

For $q \in ]0,1[$, let $\varepsilon > 0$, there exists $\delta = \left(\dfrac{\varepsilon}{M_1}\right)^{1/(\alpha q)} > 0$, such that for all $N \in \mathbf{N}^* \setminus \{1\}$, $p, p' \in A \cap \mathbf{R}^N$, $d_{1,N}(p, p') < \delta$ implies

$$\left|\frac{S_q(p)-S_q(p')}{(S_q)_{max}}\right| = \frac{\left|\sum_{i=1}^{N}\left(p_i^q-(p_i')^q\right)\right|}{|1-N^{1-q}|} \leq \frac{\sum_{i=1}^{N}\left|p_i^q-(p_i')^q\right|}{|1-N^{1-q}|} \leq \frac{N^{1-q}\left(\sum_{i=1}^{N}|p_i-p_i'|\right)^q}{|1-N^{1-q}|}$$

$$\leq \frac{N^{1-q}\left((d_{\alpha,N}(p,p'))^\alpha\right)^q}{|1-N^{1-q}|}$$

$$= \frac{N^{1-q}(d_{\alpha,N}(p,p'))^{\alpha q}}{|1-N^{1-q}|} < M_1 \delta^{\alpha q} = \varepsilon$$

For $q>1$, let $\varepsilon>0$, there exists $\delta=\left(\frac{\varepsilon}{qM_2}\right)^{1/\alpha}>0$, such that for all $N\in \mathbf{N}^*$, $p,p'\in A\cap \mathbf{R}^N$, $d_{1,N}(p,p')<\delta$ implies

$$\left|\frac{S_q(p)-S_q(p')}{(S_q)_{max}}\right| \leq \frac{\sum_{i=1}^{N}\left|p_i^q-(p_i')^q\right|}{|1-N^{1-q}|} \leq \frac{q\left(\sum_{i=1}^{N}|p_i-p_i'|\right)}{|1-N^{1-q}|}$$

$$\leq \frac{q(d_{\alpha,N}(p,p'))^\alpha}{|1-N^{1-q}|}$$

$$< qM_2\delta^\alpha = \varepsilon$$

Thus, the theorem 6 is proved.

### The $\alpha$–stability of the Incomplete $q$–expectation

**Definition 11**

Let $q\in\ ]0,+\infty[\ \backslash\{1\}$ et $C$ a element of $E_f$

The Incomplete $q$–expectation of $C$ is defined by $<C>_q(p)=\sum_{i=1}^{N}(p_i)^q C_i$ for all $p\in\left\{p\in[0,1]^N,\sum_{i=1}^{N}(p_i)^q=1\right\}$ where $N\in\mathbf{N}^*$.

**Theorem 12**

a) For $q>1$, the Incomplete $q$–expectation of $C$ is $\alpha$–stable for all $0<\alpha\leq 1$.
b) For $q\in\ ]0,1[$, the Incomplete $q$–expectation of $C$ is $\alpha$–stable for all $0<\alpha\leq q$.
c) For $q\in\ ]0,1[$, the Incomplete $q$–expectation of $C$ is $1$–unstable.

*Proof*
We have

$$\left|(<C>_q)_{N,max}\right| = \max\left(|<C>_q(p)|, p\in\left\{p\in[0,1]^N,\sum_{i=1}^{N}(p_i)^q=1\right\}\right)$$

$$= \max\left(\left|\sum_{i=1}^{N}(p_i)^q C_i\right|, \text{where } p \text{ satisfies } \sum_{i=1}^{N}(p_i)^q=1\right)$$

$$\geq \max_{i=1,\cdots,N}|C_i|\cdot 1^q = \max_{i=1,\cdots,N}|C_i|$$

a) For $q>1$, let $\varepsilon>0$, there exists $\delta=\left(\frac{\varepsilon}{q}\right)^{1/\alpha}>0$, such that for all $N\in\mathbf{N}^*$, $p,p'\in A\cap\mathbf{R}^N$, $d_{\alpha,N}(p,p')<\delta$ implies

$$\left|\frac{<C>_q(p)-<C>_q(p')}{(<C>_q)_{N,\max}}\right| \leq \frac{\sum_{i=1}^{N}\left|p_i^q-(p_i')^q\right||C_i|}{\left|(<C>_q)_{N,\max}\right|} \leq \frac{\sum_{i=1}^{N}\left|p_i^q-(p_i')^q\right||C_i|}{\max_{i=1,\cdots,N}|C_i|\cdot 1^q}$$

$$\leq \sum_{i=1}^{N}\left|p_i^q-(p_i')^q\right| \leq q\left(\sum_{i=1}^{N}|p_i-p_i'|\right)$$

$$= q(d_{1,N}(p,p')) \leq q(d_{\alpha,N}(p,p'))^\alpha < q\delta^\alpha = \varepsilon$$

b) For $q \in \,]0,1[$, let $\varepsilon > 0$, there exists $\delta = (\varepsilon)^{1/\alpha} > 0$, such that for all $N \in \mathbf{N}^*$, $p, p' \in A \cap \mathbf{R}^N$, $d_{\alpha,N}(p,p') < \delta$ implies

$$\left|\frac{<C>_q(p)-<C>_q(p')}{(<C>_q)_{N,\max}}\right| \leq \frac{\sum_{i=1}^{N}\left|p_i^q-(p_i')^q\right||C_i|}{\left|(<C>_q)_{N,\max}\right|} \leq \frac{\sum_{i=1}^{N}\left|p_i^q-(p_i')^q\right||C_i|}{\max_{i=1,\cdots,N}|C_i|\cdot 1^q}$$

$$\leq \sum_{i=1}^{N}\left|p_i^q-(p_i')^q\right| \leq \left(\sum_{i=1}^{N}|p_i-p_i'|^q\right)$$

$$= (d_{\alpha,N}(p,p'))^\alpha \leq (d_{\alpha,N}(p,p'))^\alpha < \delta^\alpha = \varepsilon$$

c) Let $q \in \,]0,1[$, and $\varepsilon = \frac{1}{2} > 0$, for all $\delta > 0$, there is $N = 2W = 2E\left(\left(\frac{2}{\delta}\right)^{\frac{q}{1-q}}\right) + 2 \in \mathbf{N}^*$, let $p, p'$ defined by

$$p = (p_i)_{i=1,\cdots,N}, \quad p_i = \left(\frac{1}{W}\right)^{1/q} \text{ if } i \text{ is odd and } p_i = 0 \text{ if } i \text{ is even}$$

$$p' = (p_i')_{i=1,\cdots,N}, \quad p_i' = \left(\frac{1}{W}\right)^{1/q} - p_i$$

We have

$$\sum_{i=1}^{N}(p_i)^q = \sum_{i=1}^{W}\left(\left(\frac{1}{W}\right)^{1/q}\right)^q = W \cdot \frac{1}{W} = 1, \quad \sum_{i=1}^{N}(p_i')^q = \sum_{i=1}^{W}\left(\left(\frac{1}{W}\right)^{1/q}\right)^q = W \cdot \frac{1}{W} = 1$$

$$d_{1,N}(p,p') = \sum_{i=1}^{N}|p_i - p_i'| = N\left(\frac{1}{W}\right)^{1/q} = 2W^{1-1/q} < \delta$$

$$C = (C_i)_{i=1,\cdots,N}, \quad C_i = 1 \text{ if } i \text{ is odd and } C_i = 0 \text{ if } i \text{ is even}$$

$$\left|(<C>_q)_{N,\max}\right| = \sum_{i=1}^{W}\left(\left(\frac{1}{W}\right)^{1/q}\right) = W^{1-1/q}$$

And $\left|\dfrac{<C>_q(p)-<C>_q(p')}{(<C>_q)_{N,\max}}\right| = \dfrac{\left|W\left(\frac{1}{W}-0\right)\right|}{W^{1-1/q}} = W^{\frac{1}{q}-1} > 1 > \varepsilon$.

Consequently, the Incomplete $q$-expectation is $1$-unstable.

**The $\alpha$-stability of the Rényi entropy**
**Definition 13**

Let $q \in \,]0,+\infty[\,\setminus\{1\}$. The Rényi entropy $R_q$ is defined on $A = \bigcup_{n \in \mathbf{N}^*}\left\{p \in [0,1]^n, \sum_{i=1}^{n}p_i = 1\right\}$ by

$$R_q(p) = \frac{\ln\left(\sum_{i=1}^{N}(p_i)^q\right)}{1-q} \text{ for all } p \in [0,1]^N.$$

**Theorem 14**
For $q \in \,]0,1[$, the Rényi entropy $R_q$ is $\alpha-$stable for all $0 < \alpha \leq q$.

*Proof.*

It's easy to see that $\left|(R_q)_{max}\right| = \left|R_q((1/N),\cdots,(1/N))\right| = \left|\frac{\ln(N^{1-q})}{1-q}\right| = \ln N$.

For $q \in \,]0,1[$, let $\varepsilon > 0$, there exists $\delta = (\varepsilon)^{1/\alpha} > 0$, such that for all $N \in \mathbf{N}^* \setminus \{1,2\}$, $p, p' \in A \cap \mathbf{R}^N$, $d_{\alpha,N}(p,p') < \delta$ implies

$$\left|\frac{R_q(p) - R_q(p')}{(R_q)_{max}}\right| = \frac{\left|\ln\left(\sum_{i=1}^{N} p_i^q\right) - \ln\left(\sum_{i=1}^{N} (p_i')^q\right)\right|}{|\ln N|}$$

By the Mean-Value Theorem, there exists $c_{q,N} \in \left(\sum_{i=1}^{N} p_i^q, \sum_{i=1}^{N} (p_i')^q\right)$ such that

$$\left|\ln\left(\sum_{i=1}^{N} p_i^q\right) - \ln\left(\sum_{i=1}^{N} (p_i')^q\right)\right| = \left|\sum_{i=1}^{N} p_i^q - \sum_{i=1}^{N} (p_i')^q\right| \frac{1}{c_{q,N}}$$

Because $\sum_{i=1}^{N} p_i^q \geq 1$ and $\sum_{i=1}^{N} (p_i')^q \geq 1$, then $\frac{1}{c_{q,N}} \leq 1$, consequently

$$\left|\frac{\ln\left(\sum_{i=1}^{N} p_i^q\right) - \ln\left(\sum_{i=1}^{N} (p_i')^q\right)}{\ln N}\right| = \frac{\left|\sum_{i=1}^{N} p_i^q - \sum_{i=1}^{N} (p_i')^q\right|}{\ln N} \frac{1}{c_{q,N}} \leq \sum_{i=1}^{N} \left|p_i^q - (p_i')^q\right|$$

By a) of lemma 10, we obtain $\left|\frac{R_q(p) - R_q(p')}{(R_q)_{max}}\right| \leq (d_{q,N}(p,p'))^q \leq (d_{\alpha,N}(p,p'))^\alpha < \delta^\alpha = \varepsilon$.

**Acknowledgements**
The authors would like to thank Prof. P. Lescot for discussions on the subject.